\documentclass[twocolumn,showpacs,superscriptaddress,aps,pra]{revtex4}
\usepackage{epsfig}
\usepackage{subfigure}
\usepackage{amssymb}
\usepackage{graphicx}
\usepackage{color}
\usepackage{amsfonts}
\usepackage{amscd}
\usepackage{hyperref}
\usepackage{amsmath}
\usepackage{epstopdf}

\begin{document}

\newcommand{\cmt}[1]{{\textcolor{red}{[#1]}}}
\newcommand{\qn}[1]{{\textcolor{red}{ (?)  #1 }}}
\newcommand{\chk}[1]{{\textcolor{green}{#1}}}
\newcommand{\del}[1]{{\textcolor{blue}{ \sout{#1}}}}
\newcommand{\rvs}[1]{{\textcolor{red}{#1}}}

\title{Quantum Zeno dynamics induced atomic entanglement in a hybrid atom-cavity-fiber system}

\author{Jiaojiao Chen}

\affiliation{Hefei Preschool Education College, Hefei 230013, China}

\author{Baolong Fang}
\affiliation{Institute of Advanced Manufacturing Engineering, Hefei University, Anhui 230022, China}

\author{Wei Xiong}

\altaffiliation{xiongweiphys@hotmail.com}
\affiliation{Institute of Advanced Manufacturing Engineering, Hefei University, Anhui 230022, China}

\date{\today }

\begin{abstract}
 Quantum entanglement is important quanum resources in quantum information sicence. Here we propose an approach to {preparing} atomic quantum entanglement in a hybrid atom-cavity-fiber system. Using quantum Zeno dynamics method, the system evolution states are always split into a series of  Zeno invariant {subspaces} consisting of dark and bright states. By choosing the initial state of the system, the bright states are all neglected and only dark states are kept to build the effective Hamiltonian. By tuning the system parameters, two-atom multiple-dimensional entanglement, two-atom Bell state, and three-atom GHZ state can be realized by one step. Our proposal provides a way to perform  quantum information processing with dark states.
 
\end{abstract}


\maketitle
\section{introduction}
Quantum entanglement, one of the most intriguing aspects of quantum mechanics,
is an essential resource to quantum information science \cite{Horodecki2009,Giraldi2001}. Therefore, generating a variety of entangled states have been paid much attention in diverse quantum systems \cite{Chuang1995,Reina2000,Zheng2000,Unanyan2003}. {For example, GHZ states, first proposed by Greenberger,
Horne, and Zeilinger, provide much stronger refutations of
local realism and offer possibilities to test quantum mechanics
against local hidden theory without inequality \cite{Greenberger90}, and it has been demonstrated that violations of local realism
by two entangled high-dimensional systems are stronger than
that by two-dimensional systems \cite{Kaszlikowski00}. Thus, much interest has
been focused on the generation of high-dimensional entangled
states.} Among various quantum systems, the emerging hybrid quantum systems, with the goal of harnessing the advantages of different subsystems to better explore new phenomena and potentially bring about novel quantum technologies (see \cite{Xiang2013,Kurizki2015} for a review), can have versatile applications in quantum information, especially for the atom-cavity-fiber hybrid systems. Because photons as flying qubits can transfer quantum information fast and atoms as stationary qubits can store quantum information for {a} long time. In {the hybrid atom-cavity-fiber} system, highly reliable swap and entangling gates are investigated via resonant interaction \cite{Serafini2006}. Then a scheme for deterministically realization of a highly reliable quantum swap gate is proposed \cite{Yin2007}. Later, a resonant interaction scheme for implementing quantum phase gates \cite{Chen2015} and a conditional phase gate \cite{Zheng2009} via a nonresonant interaction are studied. More recently, a remote swap gate through virtual photon and virtual atomic excitation is achieved \cite{Wu2019}. Also, proposals of Fredkin Gates and high-dimensional entangled states have been proposed \cite{Sun20}. Experimentally, dressed states of two separated atoms with the atom-cavity-fiber system have been observed \cite{Kato19}.
 
On the other hand, quantum Zeno effect has attracted extensive interest because it hinders the evolution out of the initial state after performing frequent measurements. In 1988, testing the Zeno effect on oscillating (mainly two-level) systems is theoretically proposed \cite{Cook1988} and demonstrated experimentally \cite{Itano1990} a few years later.  In fact, a quantum Zeno evolution does not
necessarily freeze the dynamics. By frequently projecting onto
a multidimensional subspace, the system will evolve away
from its initial state, although it remains in the “Zeno subspace”
defined by the measurement \cite{Facchi2000,Facchi2001}. Moreover, the quantum
Zeno effect can be regarded as a continuous measurement \cite{Facchi2009} without employing projection operators
and nonunitary dynamics. Since then, lots of schemes for preparation of
quantum states \cite{Luis2001,Wang2008,Li2011,Shen2011} and quantum
gates \cite{Beige2000,Franson2004,Shao2009,Franson2007} via quantum Zeno dynamics have been proposed.

Based on these advanced works, we propose a scheme based on quantum Zeno dynamics to realize two-atom multiple-dimensional entangled states, two-atom Bell state and three-atom GHZ state in a hybrid atom-cavity-fiber system. In this setup, the six-level atom is trapped in one cavity and {the} other two atoms with three-level configuration are trapped in the other cavity. Two cavities are connected by a fiber. We find that the six-level atom and one of the three-level atoms can be in non-maximally three-dimensional entangled states depending on the initial state of the six-level atom. When the coupling between the atom and the cavity is comparable to the coupling between the fiber and cavity modes, the maximally three-dimensional quantum entanglement for two separated atoms can be generated.
Also, the Bell state for the six-level atom and one of the three-level atoms can be obtained when the coupling between the atom and the cavity is much smaller than the coupling between the fiber and cavity modes, which indicates that not very strong atom-cavity coupling is enough for producing Bell states. Preparing the six-level atoms in a specially initial state, a six-demensional entangled state and GHZ state for three atoms can be achieved. Our setup may provide a way to realize more complicated quantum information processing.

The paper is organized as follows. Section \ref{sec2} introduces the physical model of our hybrid system and the method of the quantum Zeno dynamics. In Sec.~\ref{sec3}, we give a detailed description of producing various entangled states. Finally, we give a short conclusion in Sec.~\ref{sec4}.
\section{Model and quantum Zeno dynamics }\label{sec2}
\subsection{Physical model}
\begin{figure}
	\includegraphics[scale=0.5]{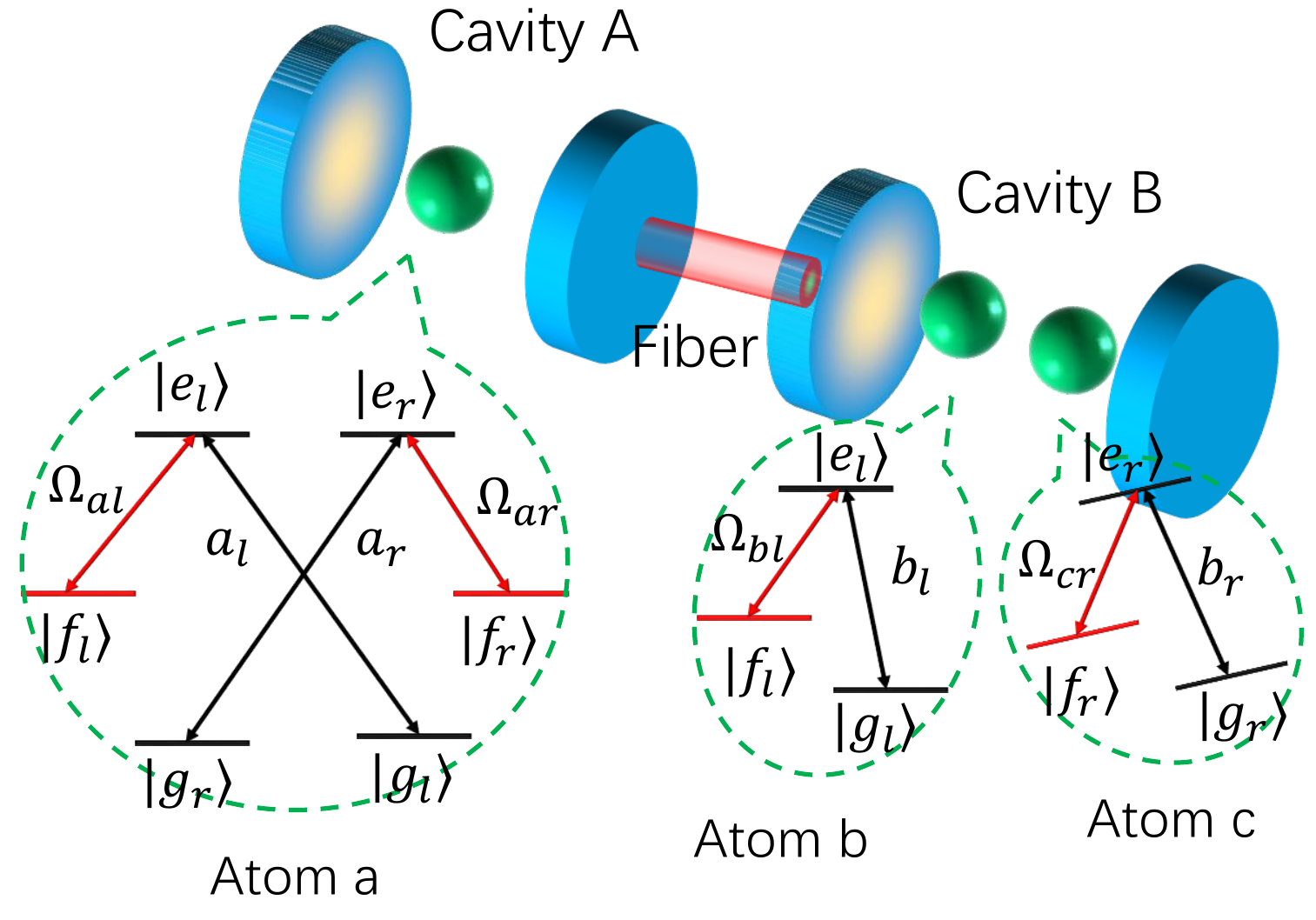}\\
	\caption{(Color online) Schematic diagram of the hybrid atom-cavity-fiber system. The atom a with six-level structure is trapped in the cavity A, and atoms b and c with three-level structure are trapped in the cavity B. Two cavities are connected by an optical fiber.}
\end{figure}
We consider a hybrid quantum system consisting of two optical cavities (labeled as ${A}$ and ${B}$) connected by an optical fiber and three atoms (labeled as atom ${a},~{b},~{c}$) with different level structures (see Fig. 1), where the atom $a$ is trapped in the cavity $A$ and the atoms $b$ and $c$ are trapped in the cavity $B$. The transition between  the level $|f_i\rangle_j$ and $|e_i\rangle_j$ is driven by a driving field with Rabi frequency $\Omega_{ij}$, where $i=l,r$ and $j=a,b,c$. The transition between $|g_i\rangle_a$ and $|e_i\rangle_a$ of the atom $a$  is coupled to the $i$ mode of the cavity $A$, and the transition between $|g_i\rangle_{b(c)}$ and $|e_i\rangle_{b(c)}$ of the atom $b~(c)$  is coupled to the $i$ mode of the cavity $B$. In the short fiber regime, i.e., $2L\bar{\nu}/2\pi c\leq 1$, the $i$ mode of the fiber can be only coupled to the $i$-mode of the cavities $A$ and $B$. Here, $L$ is the length of the optical fiber, $\bar{\nu}$ is the decay rate of the cavity fields flowing into a continuum of the fiber modes, and $c$ is the speed of light in vacuum. In the interaction representation, the total Hamitlonian of the hybrid quantum system within the rotating wave approximation can be governed by (setting $\hbar=1$)
\begin{align}
H_{tot}=H_{ac}+H_{cf}+H_d,\label{total}
\end{align}
where
\begin{align}
H_{ac}=&g_{al}|e_l\rangle_{aa}\langle g_l| c_{Al}+g_{ar}|e_r\rangle_{aa}\langle g_r| c_{Ar}\notag\\
&+g_{bl}|e_l\rangle_{bb}\langle g_l| c_{Bl}+g_{cr}|e_r\rangle_{cc}\langle g_r| c_{Br}+{\rm H.c.},\notag\\
H_{cf}=&\lambda_l d_l^\dag(c_{Al}+c_{Bl})+\lambda_r d_r^\dag(c_{Ar}+c_{Br})+{\rm H.c.},\notag\\
H_d=&\Omega_{al}|e_l\rangle _{aa}\langle f_l|+\Omega_{ar}|e_r\rangle _{aa}\langle f_r|+\Omega_{bl}|e_l\rangle _{bb}\langle f_l|\notag\\
&+\Omega_{cr}|e_r\rangle _{cc}\langle f_r|+{\rm H.c.},\label{eq2}
\end{align} 
where the first term $H_{ac}$ in Eq.~(\ref{total}) describes the interaction between the atoms and cavities with coupling strengths $g_{al},~g_{bl}$ and $g_{cr}$. The second term $H_{cf}$ in Eq.~(\ref{total}) describes the coupling between the $i$ mode of the optical fiber and the $i$ mode of the cavities $A$ and$B$ with coupling strength $\lambda_i$. The last term $H_d$ in  Eq.~(\ref{total}) describes the interaction between the atom and the driving fields. The operators $c_{Ai}$ and $c_{Bi}$ are the annihilation operators of the $i$ mode of the cavities $A$ and $B$, and ${\rm H.c.}$ in Eq.~(\ref{eq2}) denotes the hermitian conjugate.

From Eq.~(\ref{total}), atom $a$ initially prepared in the state $|f_i\rangle_a$ will be excited to the state $|e_i\rangle_a$, then a photon arising from $i$ mode is produced due to the coupling between the $i$ mode of the cavity $A$ and the transition $|e_i\rangle_a\leftrightarrow|g_i\rangle _a$ of the atom $a$. The generated photon can flow into the cavity $B$ via the interaction between the fiber and the cavities $A$ or $B$. Once the photon generated from the cavity $A$ enters the cavity $B$, it will excite the state $|g_l\rangle$ to $|e_l\rangle$ of the atom $b$ or the state $|g_r\rangle$ to $|e_r\rangle$ of the atom $c$, and vice versa.
\subsection{Quantum Zeno dynamics}
Now we briefly give a review about the quantum Zeno dynamcis induced by continuous coupling. We suppose a dynamical evolution process is governed by
\begin{align}
H_K=H_S+KH_C,
\end{align}
where $H_S$ is the Hamiltonian of the subsystem to be investigated and $H_C$ is an additional interaction Hamiltonian which performs the measurement, $K$ is the coupling strength. For a large $K$, i.e., $K\rightarrow\infty$, the subsystem of interest is dominated by the following evolution operator
\begin{align}
\mathcal{V}(t)=\lim_{K\rightarrow\infty}\exp(iKH_Ct)V(t).
\end{align}
Equivalently,
\begin{align}
\mathcal{V}(t)=\exp(-iH_Zt)
\end{align}
with
\begin{align}
H_Z=\sum_n P_n H_S P_n
\end{align}
being the Zeno Hamiltonian, in which $P_n$ is  the eigenprojection of the $H_C$ belonging to the corresponding eigenvalue $E_n$,
\begin{align}
	H_C=\sum_n E_nP_n.
\end{align}
Thus the whole system is governed by the limiting evolution operator
\begin{align}
	V_K(t)=\exp(-i H_{\rm eff} t)
\end{align} 
with
\begin{align}
	H_{\rm eff}=\sum_n KE_nP_nt+P_n H_S P_n.
\end{align}
This equation is the important result governing our following works.

\section{The dynamics of the hybrid quantum system}\label{sec3}
To begin our task, we for convenience assume $g_{al}=g_{ar}=g_{bl}=g_{cr}=g$, $\Omega_{al}=\Omega_{ar}=\Omega_1$, $\Omega_{bl}=\Omega_2$, $\Omega_{cr}=\Omega_3$, $\lambda_l=\lambda_r=\lambda$. Also, the single-photon excitation approximation is assumed in the process of the dynamics evolution.
\subsection{Initial state $|f_l\rangle_a|g_l\rangle_b|g_r\rangle_c|0_l 0_r\rangle_A|0_l 0_r\rangle_B|0_l 0_r\rangle_F$}\label{A}
Based on above analysis, we initially prepare the hybrid system in the state
\begin{align}
	|\psi_0\rangle=|f_l\rangle_a|g_l\rangle_b|g_r\rangle_c|0_l 0_r\rangle_A|0_l 0_r\rangle_B|0_l 0_r\rangle_F,
\end{align}
which indicates that the atom $a$ is initially prepared in the state $|f_l\rangle_a$, both the atoms $b$ and $c$ are respectively in the states $|g_l\rangle_b$ and $|g_r\rangle_c$, the cavities $A$, $B$ and the fiber are all in their vacuum states. Under the interaction of the Hamiltonian (\ref{total}), seven states are generated constructing a evolution subspace spanned by
\begin{align}
	U_1=\bigg\{|\phi_0\rangle,|\phi_1\rangle,|\phi_2\rangle,|\phi_3\rangle,|\phi_4\rangle,|\phi_5\rangle,|\phi_6\rangle\bigg\},\label{eq4}
\end{align}
with
\begin{align}
|\phi_0\rangle=&|\psi_0\rangle,\notag\\
|\phi_1\rangle=&|e_l\rangle_a|g_l\rangle_b|g_r\rangle_c|0_l 0_r\rangle_A|0_l 0_r\rangle_B|0_l 0_r\rangle_F,\notag\\
|\phi_2\rangle=&|g_l\rangle_a|g_l\rangle_b|g_r\rangle_c|1_l 0_r\rangle_A|0_l 0_r\rangle_B|0_l 0_r\rangle_F,\notag\\
|\phi_3\rangle=&|g_l\rangle_a|g_l\rangle_b|g_r\rangle_c|0_l 0_r\rangle_A|0_l 0_r\rangle_B|1_l 0_r\rangle_F,\notag\\
|\phi_4\rangle=&|g_l\rangle_a|g_l\rangle_b|g_r\rangle_c|0_l 0_r\rangle_A|1_l 0_r\rangle_B|0_l 0_r\rangle_F,\notag\\
|\phi_5\rangle=&|g_l\rangle_a|e_l\rangle_b|g_r\rangle_c|0_l 0_r\rangle_A|0_l 0_r\rangle_B|0_l 0_r\rangle_F,\notag\\
|\phi_6\rangle=&|g_l\rangle_a|f_l\rangle_b|g_r\rangle_c|0_l 0_r\rangle_A|0_l 0_r\rangle_B|0_l 0_r\rangle_F.\notag
\end{align}
Using these basis, the interaction Hamiltonian (\ref{total}) can be rewritten as
\begin{align}
	H_1=\begin{pmatrix}
	0 & \Omega_1 & 0 & 0 & 0 & 0 & 0\\
	\Omega_1 & 0 & g & 0 & 0 & 0 & 0\\
	0 & g & 0 & \lambda & 0 & 0 & 0\\
	0 & 0 & \lambda & 0 & \lambda & 0 & 0\\
	0 & 0 & 0 & \lambda & 0 & g & 0\\
	0 & 0 & 0 & 0 & g & 0 & \Omega_2\\
	0 & 0 & 0 & 0 & 0 & \Omega_2 & 0\\	
	\end{pmatrix}.\label{h1}
\end{align}
The eigenvectors of Hamiltonian (\ref{h1}) within the asuumption of $\{\Omega_1,\Omega_2\}\ll\{g,\lambda\}$, consisting of dark states ($|D_0\rangle,~D_1\rangle,~|D_2\rangle$) with zero eigenvalues and bright states ($|B_0\rangle,~B_1\rangle,~|B_2\rangle,~|B_3\rangle$) with non-zero values, are
\begin{align}
|D_0\rangle=&|\phi_0\rangle,~~|D_1\rangle=|\phi_6\rangle,\notag\\
|D_2\rangle=&\frac{\lambda}{g\chi}(|\phi_1\rangle-\frac{g}{\lambda}|\phi_3\rangle+|\phi_5\rangle),\notag\\
|B_0\rangle=&\frac{1}{2}(-|\phi_1\rangle-|\phi_2\rangle+|\phi_4\rangle+|\phi_5\rangle),\notag\\
|B_1\rangle=&\frac{1}{2}(-|\phi_1\rangle+|\phi_2\rangle-|\phi_4\rangle+|\phi_5\rangle),\notag\\
|B_2\rangle=&2\chi(|\phi_1\rangle+\chi|\phi_2\rangle+\frac{\lambda}{g}|\phi_3\rangle-\chi|\phi_4\rangle+|\phi_5\rangle),\notag\\
|B_3\rangle=&2\chi(|\phi_1\rangle-\chi|\phi_2\rangle+\frac{\lambda}{g}|\phi_3\rangle-\chi|\phi_4\rangle+|\phi_5\rangle),\label{eq6}
\end{align}
where $\chi=(1+2\lambda^2/g^2)^{\frac{1}{2}}$ and the correspinding eigenvalues are $E_0=0,~E_1=0,~E_2=0,~E_3=g,~E_4=-g,~E_5=g\chi,~E_6=-g\chi$. Note that the whole Hilbert subspace for Hamiltonian (\ref{h1}) under $\{\Omega_1,\Omega_2\}\ll\{g,\lambda\}$ can be divided into five invariant subspaces as following,
\begin{align}
\Gamma_{p_0}=&\{|D_0\rangle,|D_1\rangle,|D_2\rangle\},~\Gamma_{p_1}=\{|B_0\rangle\}\notag\\
\Gamma_{p_2}=&\{|B_1\rangle\},~\Gamma_{p_3}=\{|B_2\rangle\},~\Gamma_{p_4}=\{|B_3\rangle\}.
\end{align}
Here $p_j=|\alpha\rangle\langle \alpha |$ with $j=0,1,2,3,4$ is the projector and $\alpha \in\{\Gamma_{p_0},\Gamma_{p_1},\Gamma_{p_2},\Gamma_{p_3},\Gamma_{p_4}\}$. According to quantum Zeno dynamics method, the Hamiltonian in Eq.~(\ref{h1}) reduces to
\begin{align}
	\mathcal{H}_1=&\sum_{j=0} E_j p_j+p_j H_d p_j\notag\\
	=&g\big[|B_0\rangle \langle B_0|-|B_1\rangle \langle B_1|+\chi(|B_2\rangle \langle B_2|-|B_3\rangle \langle B_3|)\big]\notag\\
	&+\frac{\lambda}{g\chi}(\Omega_1|D_0\rangle\langle D_2|+\Omega_2|D_1\rangle\langle D_2|+{\rm H.c.}).\label{eq8}
\end{align}
Since the system is initially prepared in the state $|D_0\rangle$, the system will always evolve in the Zeno subspace spanned by$\{|D_0\rangle,|D_2\rangle,|B_2\rangle\}$. The corresponding dynamics can be governed by the effective Hamiltonian
\begin{align}
H_{\rm eff}=\frac{\lambda}{g\chi}(\Omega_1|D_0\rangle\langle D_2|+\Omega_2|D_1\rangle\langle D_2|+{\rm H.c.}),
\end{align}
and at time $\tau$, the system state is
\begin{align}
|\psi_\tau\rangle=e^{-i H_{\rm eff}\tau}|\psi_0\rangle=A_1|D_0\rangle+A_2|D_2\rangle+A_3|D_1\rangle,
\label{eq10}
\end{align}
where
\begin{align}
	A_1=&\sin^2\theta+\cos^2\theta \cos(\Omega\lambda \tau/g\chi),\notag\\
	A_2=&-i \cos\theta\sin(\Omega\lambda \tau/g\chi)\notag\\
	A_3=&\frac{1}{2}\sin(2\theta)\big[\cos(\Omega\lambda \tau/g\chi)-1\big]
\end{align}
with $\Omega_1^2+\Omega_2^2=\Omega^2$ and $\tan\theta=\Omega_2/\Omega_1$. By controlling the system parameters $\{\Omega_1,\Omega_2,\lambda,g,\tau\}$, different quantum information task can be achieved. For example, when $\Omega_2\ll\Omega_1$, $\sin\theta\rightarrow 0$ and $\cos\theta\rightarrow 1$. Thus, $A_1\rightarrow \cos(\Omega\lambda \tau/g\chi)$, $A_2\rightarrow -i \sin(\Omega\lambda \tau/g\chi)$ and $A_3\rightarrow 0$. Then we choose $\Omega\lambda \tau/g\chi=(2k-1)\pi/2$ with $k=1,2,3,...$, leading to $\cos(\Omega\lambda \tau/g\chi)=0$ and $\cos(\Omega\lambda \tau/g\chi)=1$. This condition can be achieved by tuning the parameters $\{\Omega_1,\Omega_2,\lambda,g,\tau\}$, which greatly reduces the difficulties   experimentally. With this condition, Eq. (\ref{eq10}) becomes
\begin{align}
	|\psi_\tau\rangle_{\rm ST}=-i|D_2\rangle,\label{eq12}
\end{align}
which indicates quantum state transition between two degenerate states is allowed and the state $|D_1\rangle$ is safely neglected. Note that when the photons in the $l$ mode of the optical fiber experience a Hadamard gate, i.e., $|0\rangle\rightarrow\frac{1}{\sqrt{2}}(|0\rangle+|1\rangle)$ and $|1\rangle\rightarrow\frac{1}{\sqrt{2}}(|0\rangle-|1\rangle)$, and then trace the degrees of the freedom of the atom $b$ and the photons in the cavity and fiber modes out, the quantum state in Eq. (\ref{eq12}) further reduces to
\begin{align}
|\psi_\tau\rangle_{\rm TD}=\frac{1}{\sqrt{3}}(|e_l\rangle |g_l\rangle-|g_l\rangle |g_l\rangle+|g_l\rangle |e_l\rangle)_{ab}\label{eq13}
\end{align}
by setting $g=\lambda$, which is a three-dimensional quantum entanglement between atoms $a$ and $b$. However, when $g\ll\lambda$, Eq. (\ref{eq12}) describes a Bell state between atoms $a$ and $b$, i.e.,
\begin{align}
|\psi_\tau\rangle_{\rm Bell}=\frac{1}{\sqrt{2}}(|e_l\rangle |g_l\rangle+|g_l\rangle |e_l\rangle)_{ab}.\label{eq14}
\end{align}
On the other hand, Eq.~(\ref{eq10}) can be rewritten as
\begin{align}
	|\psi_\tau\rangle_{\rm SW}=|D_1\rangle\label{eq15}
\end{align}
by setting $\Omega_1=\Omega_2=\Omega/\sqrt{2}$ and $\Omega\lambda \tau/g\chi=k\pi$ with $k=1,2,3,...$. This implies that state swapping ($|f_l\rangle_a|g_l\rangle_b \leftrightarrow |g_l\rangle_a|f_l\rangle_b$) between atoms $a$ and $b$ is achieved.

\subsection{Initial state $|f_r\rangle_a|g_l\rangle_b|g_r\rangle_c|0_l 0_r\rangle_A|0_l 0_r\rangle_B|0_l 0_r\rangle_F$}\label{B}

Below we assume that the initial state of the atom-cavity-fiber hybrid system is
\begin{align}
|\psi_0^\prime\rangle=|f_r\rangle_a|g_l\rangle_b|g_r\rangle_c|0_l 0_r\rangle_A|0_l 0_r\rangle_B|0_l 0_r\rangle_F,
\end{align}
which indicates that the atom $a$ is initially prepared in the state $|f_r\rangle_a$, both the atoms $b$ and $c$ are respectively in the states $|g_l\rangle_b$ and $|g_r\rangle_c$, the cavities $A$, $B$ and the fiber are all in their vacuum states. Under the interaction of the Hamiltonian (\ref{total}), seven states are generated constructing a evolution subspace spanned by
\begin{align}
U_1^\prime=\bigg\{|\phi_0^\prime\rangle,|\phi_1^\prime\rangle,|\phi_2^\prime\rangle,|\phi_3^\prime\rangle,|\phi_4^\prime\rangle,|\phi_5^\prime\rangle,|\phi_6^\prime\rangle\bigg\},\label{eq17}
\end{align}
with
\begin{align}
|\phi_0^\prime\rangle=&|\psi_0^\prime\rangle,\notag\\
|\phi_1^\prime\rangle=&|e_r\rangle_a|g_l\rangle_b|g_r\rangle_c|0_l 0_r\rangle_A|0_l 0_r\rangle_B|0_l 0_r\rangle_F,\notag\\
|\phi_2^\prime\rangle=&|g_r\rangle_a|g_l\rangle_b|g_r\rangle_c|1_l 0_r\rangle_A|0_l 0_r\rangle_B|0_l 0_r\rangle_F,\notag\\
|\phi_3^\prime\rangle=&|g_r\rangle_a|g_l\rangle_b|g_r\rangle_c|0_l 0_r\rangle_A|0_l 0_r\rangle_B|0_l 1_r\rangle_F,\notag\\
|\phi_4^\prime\rangle=&|g_r\rangle_a|g_l\rangle_b|g_r\rangle_c|0_l 0_r\rangle_A|0_l 1_r\rangle_B|0_l 0_r\rangle_F,\notag\\
|\phi_5^\prime\rangle=&|g_r\rangle_a|g_l\rangle_b|e_r\rangle_c|0_l 0_r\rangle_A|0_l 0_r\rangle_B|0_l 0_r\rangle_F,\notag\\
|\phi_6^\prime\rangle=&|g_r\rangle_a|g_l\rangle_b|f_r\rangle_c|0_l 0_r\rangle_A|0_l 0_r\rangle_B|0_l 0_r\rangle_F.\notag
\end{align}
Using these basis, the interaction Hamiltonian (\ref{total}) can be rewritten as
\begin{align}
H_1^\prime=\begin{pmatrix}
0 & \Omega_1 & 0 & 0 & 0 & 0 & 0\\
\Omega_1 & 0 & g & 0 & 0 & 0 & 0\\
0 & g & 0 & \lambda & 0 & 0 & 0\\
0 & 0 & \lambda & 0 & \lambda & 0 & 0\\
0 & 0 & 0 & \lambda & 0 & g & 0\\
0 & 0 & 0 & 0 & g & 0 & \Omega_3\\
0 & 0 & 0 & 0 & 0 & \Omega_3 & 0\\	
\end{pmatrix}.\label{eq18}
\end{align}
The eigenvectors of Hamiltonian (\ref{eq18}) within the asuumption of $\{\Omega_1,\Omega_3\}\ll\{g,\lambda\}$, consisting of dark states ($|D_0^\prime\rangle,~D_1^\prime\rangle,~|D_2^\prime\rangle$) with zero eigenvalues and bright states ($|B_0^\prime\rangle,~B_1^\prime\rangle,~|B_2^\prime\rangle,~|B_3^\prime\rangle$) with non-zero values, are
\begin{align}
|D_0^\prime\rangle=&|\phi_0^\prime\rangle,~~|D_1^\prime\rangle=|\phi_6^\prime\rangle,\notag\\
|D_2^\prime\rangle=&\frac{\lambda}{g\chi}(|\phi_1^\prime\rangle-\frac{g}{\lambda}|\phi_3^\prime\rangle+|\phi_5^\prime\rangle),\notag\\
|B_0^\prime\rangle=&\frac{1}{2}(-|\phi_1^\prime\rangle-|\phi_2^\prime\rangle+|\phi_4^\prime\rangle+|\phi_5^\prime\rangle),\notag\\
|B_1^\prime\rangle=&\frac{1}{2}(-|\phi_1^\prime\rangle+|\phi_2^\prime\rangle-|\phi_4^\prime\rangle+|\phi_5^\prime\rangle),\notag\\
|B_2^\prime\rangle=&2\chi(|\phi_1^\prime\rangle+\chi|\phi_2^\prime\rangle+\frac{\lambda}{g}|\phi_3^\prime\rangle-\chi|\phi_4^\prime\rangle+|\phi_5^\prime\rangle),\notag\\
|B_3^\prime\rangle=&2\chi(|\phi_1^\prime\rangle-\chi|\phi_2^\prime\rangle+\frac{\lambda}{g}|\phi_3^\prime\rangle-\chi|\phi_4^\prime\rangle+|\phi_5^\prime\rangle),\label{eq19}
\end{align}
where $\chi=(1+2\lambda^2/g^2)^{\frac{1}{2}}$ and the correspinding eigenvalues are $E_0^\prime=0,~E_1^\prime=0,~E_2^\prime=0,~E_3^\prime=g,~E_4^\prime=-g,~E_5^\prime=g\chi,~E_6^\prime=-g\chi$. Note that the whole Hilbert subspace for Hamiltonian (\ref{eq18}) under $\{\Omega_1,\Omega_3\}\ll\{g,\lambda\}$ can be divided into five invariant subspaces as following,
\begin{align}
\Gamma_{p_0^\prime}=&\{|D_0^\prime\rangle,|D_1^\prime\rangle,|D_2^\prime\rangle\},~\Gamma_{p_1^\prime}=\{|B_0^\prime\rangle\}\notag\\
\Gamma_{p_2^\prime}=&\{|B_1^\prime\rangle\},~\Gamma_{p_3^\prime}=\{|B_2^\prime\rangle\},~\Gamma_{p_4^\prime}=\{|B_3^\prime\rangle\}.
\end{align}
Here $p_j^\prime=|\alpha^\prime\rangle\langle \alpha^\prime |$ with $j=0,1,2,3,4$ is the projector and $\alpha^\prime \in\{\Gamma_{p_0^\prime},\Gamma_{p_1^\prime},\Gamma_{p_2^\prime},\Gamma_{p_3^\prime},\Gamma_{p_4^\prime}\}$. According to quantum Zeno dynamics method, the Hamiltonian in Eq.~(\ref{eq18}) reduces to
\begin{align}
\mathcal{H}_1^\prime=&\sum_{j=0} E_j^\prime p_j^\prime+p_j^\prime H_d p_j^\prime\notag\\
=&g\big[|B_0^\prime\rangle \langle B_0^\prime|-|B_1^\prime\rangle \langle B_1^\prime|+\chi(|B_2^\prime\rangle \langle B_2^\prime|-|B_3^\prime\rangle \langle B_3^\prime|)\big]\notag\\
&+\frac{\lambda}{g\chi}(\Omega_1|D_0^\prime\rangle\langle D_2^\prime|+\Omega_3|D_1^\prime\rangle\langle D_2^\prime|+{\rm H.c.}).\label{eq21}
\end{align}
Since the system is initially prepared in the state $|D_0^\prime\rangle$, the system will always evolve in the Zeno subspace spanned by$\{|D_0^\prime\rangle,|D_2^\prime\rangle,|B_2^\prime\rangle\}$. The corresponding dynamics can be governed by the effective Hamiltonian
\begin{align}
H_{\rm eff}^\prime=\frac{\lambda}{g\chi}(\Omega_1|D_0^\prime\rangle\langle D_2^\prime|+\Omega_3|D_1^\prime\rangle\langle D_2^\prime|+{\rm H.c.}),
\end{align}
and at time $\tau^\prime$, the system state is
\begin{align}
|\psi_{\tau^\prime}^\prime\rangle=e^{-i H_{\rm eff}^\prime\tau^\prime}|\psi_0^\prime\rangle=A_1^\prime|D_0^\prime\rangle+A_2^\prime|D_2^\prime\rangle+A_3^\prime|D_1^\prime\rangle,
\label{eq23}
\end{align}
where
\begin{align}
A_1^\prime=&\sin^2\theta^\prime+\cos^2\theta^\prime \cos(\bar{\Omega}\lambda \tau^\prime/g\chi),\notag\\
A_2^\prime=&-i \cos\theta^\prime\sin(\bar{\Omega}\lambda \tau^\prime/g\chi)\notag\\
A_3^\prime=&\frac{1}{2}\sin(2\theta^\prime)\big[\cos(\bar{\Omega}\lambda \tau^\prime/g\chi)-1\big]
\end{align}
with $\Omega_1^2+\Omega_3^2=\bar{\Omega}^{2}$ and $\tan\theta^\prime=\Omega_3/\Omega_1$. 

By tuning the system parameters $\{\Omega_1,\Omega_3,\lambda,g,\tau^\prime\}$, the similar quantum information tasks can be achieved in Sec.~\ref{A}. For example, when $\Omega_3\ll\Omega_1$, $\sin\theta^\prime\rightarrow 0$ and $\cos\theta^\prime\rightarrow 1$. Thus, $A_1^\prime\rightarrow \cos(\bar{\Omega}\lambda \tau^\prime/g\chi)$, $A_2^\prime\rightarrow -i \sin(\bar{\Omega}\lambda \tau^\prime/g\chi)$ and $A_3^\prime\rightarrow 0$. Then we choose $\bar{\Omega}\lambda \tau^\prime/g\chi=(2k-1)\pi/2$ with $k=1,2,3,...$, leading to $\cos(\bar{\Omega}\lambda \tau^\prime/g\chi)=0$ and $\cos(\bar{\Omega}\lambda \tau^\prime/g\chi)=1$. This condition can be achieved by tuning the parameters $\{\Omega_1,\Omega_3,\lambda,g,\tau^\prime\}$, which greatly reduces the difficulties experimentally. With this condition, Eq. (\ref{eq23}) becomes
\begin{align}
|\psi_{\tau^\prime}^\prime\rangle_{\rm ST}=-i|D_2^\prime\rangle,\label{eq25}
\end{align}
which indicates quantum state transition between two degenerate states is allowed and the state $|D_1^\prime\rangle$ is safely neglected. Note that when the photons in the $r$ mode of the optical fiber experience a Hadamard gate, i.e., $|0_r\rangle\rightarrow\frac{1}{\sqrt{2}}(|0_r\rangle+|1_r\rangle)$ and $|1_r\rangle\rightarrow\frac{1}{\sqrt{2}}(|0_r\rangle-|1_r\rangle)$, and then trace the degrees of the freedom of the atom $b$ and the photons in the cavity and fiber modes out, the quantum state in Eq. (\ref{eq25}) further reduces to
\begin{align}
|\psi_{\tau^\prime}^\prime\rangle_{\rm TD}=\frac{1}{\sqrt{3}}(|e_r\rangle |g_r\rangle-|g_r\rangle |g_r\rangle+|g_r\rangle |e_r\rangle)_{ac}\label{eq26}
\end{align}
by setting $g=\lambda$, which is a three-dimensional quantum entanglement between atoms $a$ and $c$. However, when $g\ll\lambda$, Eq. (\ref{eq25}) describes a Bell state between atoms $a$ and $c$, i.e.,
\begin{align}
|\psi_{\tau^\prime}^\prime\rangle_{\rm Bell}=\frac{1}{\sqrt{2}}(|e_r\rangle |g_r\rangle+|g_r\rangle |e_r\rangle)_{ac}.\label{eq27}
\end{align}
However, by setting $\Omega_1=\Omega_3=\bar{\Omega}/\sqrt{2}$ and $\bar{\Omega}\lambda \tau^\prime/g\chi=k\pi$ with $k=1,2,3,...$, Eq.~(\ref{eq23}) becomes
\begin{align}
|\psi_{\tau^\prime}^\prime\rangle_{\rm SW}=|D_1^\prime\rangle.\label{eq28}
\end{align}
This implies that state swapping between atoms $a$ and $c$, i.e., $|f_r\rangle_a|g_r\rangle_c \leftrightarrow |g_r\rangle_a|f_r\rangle_c$, can be achieved.

\subsection{Initial state $\Psi_0=|\psi_0\rangle+|\psi_0^\prime\rangle$}

Based on the analysis in Sec.~\ref{A} and Sec.~\ref{B}, we further consider that the atom $a$ is initially prepared in the superposition state $|g_l\rangle_a+|e_r\rangle_a$, atoms $b$ and $c$ are in the states $|g_l\rangle_b$ and $|g_r\rangle_c$, respectively. Cavities $A$ and $B$, and the fiber are all in their vacuum states. Therefore, the initial state of the considered system can be written as
\begin{align}
|\Psi_0\rangle=|\psi_0\rangle+|\psi_0^\prime\rangle.
\end{align}
According to Eqs.~(\ref{eq4}) and (\ref{eq17}), the evolution space governed by Eq.~(\ref{total}) with the initial state $|\Psi_0\rangle$ is spanned by states in $U_1$ plus $U_1^\prime$, i.e.,
\begin{align}
U=\{|\Phi_0\rangle,|\Phi_1\rangle,|\Phi_2\rangle,|\Phi_3\rangle,|\Phi_4\rangle,|\Phi_5\rangle,|\Phi_6\rangle\},\label{eq30}
\end{align}
where
\begin{align}
|\Phi_0\rangle=&|\Psi_0\rangle,\notag\\
|\Phi_1\rangle=&|\phi_1\rangle+|\phi_1^\prime\rangle,\notag\\
|\Phi_2\rangle=&|\phi_2\rangle+|\phi_2^\prime\rangle,\notag\\
|\Phi_3\rangle=&|\phi_3\rangle+|\phi_3^\prime\rangle,\notag\\
|\Phi_4\rangle=&|\phi_4\rangle+|\phi_4^\prime\rangle,\notag\\
|\Phi_5\rangle=&|\phi_5\rangle+|\phi_5^\prime\rangle,\notag\\
|\Phi_6\rangle=&|\phi_6\rangle+|\phi_6^\prime\rangle.
\end{align}
Using the basis in Eq.~(\ref{eq30}), the system Hamiltonian in Eq.~(\ref{total}) reduces to
\begin{align}
H=
\begin{pmatrix}
0 & \Omega_1 & 0 & 0 & 0 & 0 & 0\\
\Omega_1 & 0 & g & 0 & 0 & 0 & 0\\
0 & g & 0 & \lambda & 0 & 0 & 0\\
0 & 0 & \lambda & 0 & \lambda & 0 & 0\\
0 & 0 & 0 & \lambda & 0 & g & 0\\
0 & 0 & 0 & 0 & g & 0 & \Omega_2+\Omega_3\\
0 & 0 & 0 & 0 & 0 & \Omega_2+\Omega_3 & 0\\	
\end{pmatrix}.\label{eq32}
\end{align}

When we choose $\Omega_1,\Omega_2,\Omega_3\ll g,\lambda$, the states in Eqs.~(\ref{eq6}) and (\ref{eq19}) can be simutaneously obtained. Thus, the eigenvectors of the Hamitlonian in Eq.~(\ref{total})  are
\begin{align}
|\mathcal{D}_0\rangle=&|D_0\rangle+|D_0^\prime\rangle,\notag\\
|\mathcal{D}_1\rangle=&|D_1\rangle+|D_1^\prime\rangle,\notag\\
|\mathcal{D}_2\rangle=&|D_2\rangle+|D_2^\prime\rangle,\notag\\
|\mathcal{B}_0\rangle=&|B_0\rangle+|B_0^\prime\rangle,\notag\\
|\mathcal{B}_1\rangle=&|B_1\rangle+|B_1^\prime\rangle,\notag\\
|\mathcal{B}_2\rangle=&|B_2\rangle+|B_2^\prime\rangle,\notag\\
|\mathcal{B}_3\rangle=&|B_3\rangle+|B_3^\prime\rangle,
\end{align}
with $\chi=(1+2\lambda^2/g^2)^{\frac{1}{2}}$ and the correspinding eigenvalues are $\mathcal{E}_0=0,~\mathcal{E}_1=0,~\mathcal{E}_2=0,~\mathcal{E}_3=g,~\mathcal{E}_4=-g,~\mathcal{E}_5=g\chi,~\mathcal{E}_6=-g\chi$. The space consisting of these eigenvectors can be split into five invariant Zeno subspace
\begin{align}
 \Gamma_{\mathcal{P}_0}=&\{|\mathcal{D}_0\rangle,|\mathcal{D}_1\rangle,|\mathcal{D}_2\rangle\},~\Gamma_{\mathcal{P}_1}=\{|\mathcal{B}_0\rangle\}\notag\\
\Gamma_{\mathcal{P}_2}=&\{|\mathcal{B}_1\rangle\},~\Gamma_{\mathcal{P}_3}=\{|\mathcal{B}_2\rangle\},~\Gamma_{\mathcal{P}_4}=\{|\mathcal{B}_3\rangle\}.
\end{align}
Here $\mathcal{P}_j=|\bar{\alpha}\rangle\langle \bar{\alpha}|$ with $j=0,1,2,3,4$ is the projector and $\bar{\alpha} \in\{\Gamma_{\mathcal{P}_0},\Gamma_{\mathcal{P}_1},\Gamma_{\mathcal{P}_2},\Gamma_{\mathcal{P}_3},\Gamma_{\mathcal{P}_4}\}$. Combining Eqs.~(\ref{eq8}) and (\ref{eq21}), the Hamiltonian in Eq.~(\ref{eq32}) reduces to
\begin{align}
\mathcal{H}=&\sum_{j=0} \mathcal{E}_j \mathcal{P}_j+\mathcal{P}_j H_d \mathcal{P}_j\notag\\
=&g\big[|\mathcal{B}_0\rangle \langle \mathcal{B}_0|-|\mathcal{B}_1\rangle \langle \mathcal{B}_1|+\chi(|\mathcal{B}_2\rangle \langle \mathcal{B}_2|-|\mathcal{B}_3\rangle \langle \mathcal{B}_3|)\big]\notag\\
&+\frac{\lambda}{g\chi}(\Omega_1|\mathcal{D}_0\rangle\langle \mathcal{D}_2|+(\Omega_2|D_1\rangle+\Omega_3|D_1^\prime)\rangle\langle \mathcal{D}_2|+{\rm H.c.}).\label{eq34}
\end{align}
When the system state is initially prepared in the state $|\mathcal{D}_0\rangle$, Eq.~(\ref{eq34})  can be rewritten as
\begin{align}
\mathcal{H}_{\rm eff}=\frac{\lambda}{g\chi}(\Omega_1|\mathcal{D}_0\rangle\langle \mathcal{D}_2|+(\Omega_2|D_1\rangle+\Omega_3|D_1^\prime)\rangle\langle \mathcal{D}_2|+{\rm H.c.}).
\end{align}
At time $T$, the system state is
\begin{align}
	|\Psi_T\rangle=&\mathcal{A}_1|{D}_0\rangle+\mathcal{A}_1^\prime|{D}_0^\prime\rangle+\mathcal{A}_2|{D}_2\rangle\notag\\
	&+\mathcal{A}_2^\prime|D_2^\prime\rangle+\mathcal{A}_3|D_1\rangle+\mathcal{A}_3^\prime|D_1^\prime\rangle,
\end{align}
where
\begin{align}
\mathcal{A}_1=&A_1,~\mathcal{A}_1^\prime=A_1^\prime,~\mathcal{A}_2=A_2,\notag\\
\mathcal{A}_2^\prime=&A_2^\prime,~\mathcal{A}_3=A_3,~\mathcal{A}_3^\prime=A_3^\prime.
\end{align}
Obviously, $\Omega_2,\Omega_3\ll\Omega_1$, the state $q$ becomes
\begin{align}
|\Psi_T\rangle_{\rm ST}=-i\mathcal{D}_2,
\end{align}
which indicates that quantum state transition between two dark states can be achieved. The similar process from Eq.~(\ref{eq13}) to Eq.~(\ref{eq15}) or Eq.~(\ref{eq26}) to Eq.~(\ref{eq28}) is performed, the high-dimensional quantum entangled state among three atoms with setting $g=\lambda$
\begin{align}
|\Psi_T\rangle _{\rm HD}=&\frac{1}{\sqrt{6}}[(|e_r\rangle|g_l\rangle |g_r\rangle-|g_r\rangle |g_l\rangle|g_r\rangle+|g_r\rangle|g_l\rangle |e_r\rangle)_{abc}\notag\\
&+(|e_l\rangle |g_l\rangle|g_r\rangle-|g_l\rangle |g_l\rangle|g_r\rangle+|g_l\rangle |e_l\rangle|g_r\rangle)_{abc}]
\end{align}
can be obtained, and the three atoms GHZ state 
\begin{align}
	|\Psi_T\rangle _{\rm GHZ}=\frac{1}{\sqrt{2}}\mathcal{D}_1
\end{align}
is also realized when $\Omega_1=\Omega_2=\Omega_3$ and $\Omega_3\lambda T/g\chi=k\pi$ are satisfied.

\section{conclusion}\label{sec4}

In conclusion, we have proposed a hybrid system to entangle remote atoms trapped in two cavities linked by an optical fiber. Employing quantum Zeno dynamics method, bright states of the system are all neglected and dark states are kept to build the effective Hamiltonian. By considering the relative strength between the atom-cavity coupling and cavity-fiber coupling and preparing the six-level atom in different initial state, the Bell state, the two-atom three-dimensional entangled states, the three-atom GHZ state  and six-dimentional entangled state are achieved. Our proposal provides a path to perform quantum information processing with dark states.

\section*{Acknowledgments}

This work is supported by the National Natural Science Foundation of China (Grant No. 11804074), Talent Development Funding of Hefei University (Grant No. 18-19RC60), Candidates of Hefei University Academic Leader(Grant No.2016dtr02), Major Project of Hefei Preschool Education College (Grant No.hyzzd2018003).

\end{document}